\documentclass[numsec,webpdf,contemporary,large]{oup-authoring-template}%

\graphicspath{{Fig/}}

\theoremstyle{thmstyleone}%
\theoremstyle{thmstyletwo}%
\theoremstyle{thmstylethree}%

\usepackage{booktabs}
\usepackage{graphicx}
\usepackage[caption=false,font=small]{subfig}
\usepackage{ulem}
\usepackage{xcolor}

\begin{document}

\journaltitle{Under review}
\copyrightyear{2025}
\pubyear{2025}
\appnotes{Paper}

\firstpage{1}


\title{Structure-Aware Antibody Design with Affinity-Optimized Inverse Folding}

\author[1]{Xinyan Zhao}
\author[1]{Yi-Ching Tang}
\author[2,$\dagger$]{Rivaaj Monsia}
\author[1]{Victor J Cantu}
\author[4]{Ashwin Kumar Ramesh}
\author[3]{Xiaozhong Liu}
\author[4]{Zhiqiang An}
\author[1]{Xiaoqian Jiang}
\author[1,$\ast$]{Yejin Kim}

\authormark{Author Name et al.}

\address[1]{\orgdiv{McWilliams School of Biomedical Informatics}, \orgname{University of Texas Health Science Center at Houston}, \postcode{77030}, \state{Texas}, \country{United States}}
\address[2]{\orgdiv{Department of Computer Science}, \orgname{The University of Texas at Austin}, \postcode{78712}, \state{Texas}, \country{United States}}
\address[3]{\orgdiv{Computer Science Department}, \orgname{Worcester Polytechnic Institute}, \postcode{01609}, \state{Massachusetts}, \country{United States}}
\address[4]{\orgdiv{Texas Therapeutics Institute, Brown Foundation Institute of Molecular Medicine}, \orgname{University of Texas Health Science Center at Houston}, \postcode{77030}, \state{Texas}, \country{United States}}

\corresp[$\dagger$]{Work done while at University of Texas Health Science Center.}
\corresp[$\ast$]{Corresponding author. \href{Yejin.Kim@uth.tmc.edu}{Yejin.Kim@uth.tmc.edu}}



\abstract{
\textbf{Motivation:} The clinical efficacy of antibody therapeutics critically depends on achieving high-affinity engagement with their targets, yet laboratory affinity-maturation campaigns are slow and costly. For in silico computation, most protein language models (PLMs) are not trained to favour high-affinity antibodies, and current preference optimization approaches introduce substantial computational overhead without demonstrable improvements in affinity-related performance. Therefore, this works propose SimBinder-IF, we convert the inverse folding model ESM-IF into an antibody sequence generator by freezing its structure encoder and training only its decoder to prefer experimentally stronger binders through preference optimization.\\
\textbf{Results:} On the 11-assay AbBiBench benchmark, SimBinder-IF achieves a 55\% relative improvement in mean Spearman correlation between log-likelihood scores and experimentally measured binding affinity compared to vanilla ESM-IF (increasing from 0.264 to 0.410). In zero-shot generalization tests across four unseen antigen–antibody complexes, the correlation increases by 156\% (from 0.115 to 0.294). SimBinder-IF also outperforms baselines in top-10 precision for $\geq$ 10-fold affinity improvements. These results indicate the ability of SimBinder-IF to generate antibodies with higher binding affinity compared to general PLMs. To further validate SimBinder-IF, we conducted a case study redesigning the antibody F045-092 to target A/California/04/2009 (pdmH1N1), a strain for which the original F045-092 exhibits little to no binding affinity. SimBinder-IF proposed antibody variants with markedly lower predicted binding free energy change values relative to those designed by the original ESM-IF model (mean $\Delta \Delta G$ -75.16 vs -46.57), confirming the improved affinity of the designed antibodies. Notably, the number of trainable parameters in SimBinder-IF constitutes only ~18\% of the full ESM-IF model, highlighting the parameter efficiency of SimBinder-IF for high-affinity antibody candidate generation.
}

\keywords{Antibody Binding Affinity Optimization; Inverse Folding; Protein Language Model; Preference Optimization}


\maketitle
\thispagestyle{empty}
\section{Introduction}

From the approval of the first monoclonal antibody, Muromonab‐CD3, in 1986 to the end of 2024, the FDA had cumulatively authorized over 160 monoclonal antibodies \cite{wang2025antibody}. In 2024 alone, 13 new antibodies were approved, setting a record for annual approvals \cite{de2025pharmaceutical}. The binding affinity between an antibody and its antigen is a principal determinant of an antibody’s therapeutic potential. Higher-affinity variants typically result in enhanced in vivo potency, reduced dosing frequency, consequently mitigating immunogenicity risk and decreasing the likelihood of immune escape \cite{dall2006properties, muecksch2021affinity, dyson2020beyond}. Conversely, progress with candidates that show sub-optimal affinities are frequently discontinued in pre-clinical or early clinical stages despite otherwise favorable profiles. Although phage display enables high-throughput library screening, downstream validation—such as surface plasmon resonance (SPR) or biolayer interferometry (BLI)—remains low-throughput and costly. Moreover, the method itself can introduce bias toward highly expressed or stable clones. Machine learning-based antibody design approaches complement these experimental techniques by efficiently navigating the vast search space and generating high-affinity antibody variants that may not be readily accessible through experimental approaches alone \cite{shanker2024unsupervised, pacesa2025one}. A computational model that can directly generate candidate antibody sequences with enriched affinity holds the promise of dramatically accelerating lead discovery and reducing experimental burden. 

Self-supervised protein language models (PLMs) such as the Evolutionary Scale Modeling (ESM) \cite{lin2023evolutionary, hayes2025simulating, hsu2022learning} series have demonstrated their ability to learn fundamental “protein grammar”, they are capable of capturing secondary and tertiary structural features and facilitating the inference of long-range residue–residue contacts \cite{lin2023evolutionary, zhang2024protein}. Antibodies constitute a distinct protein class with a modular $\mathrm{H_2L_2}$ architecture and discrete variable and constant domains \cite{edelman1973antibody, mcconnell2025new}. Diversity focused in six CDR loops—sculpted by somatic hypermutation and clonal selection—confers epitope-level specificity and affinity, distinguishing antibodies from most other proteins. However, naturally evolved proteins do not optimize solely for binding affinity, and most general PLMs—trained by self-supervision on large-scale protein databases—lack any explicit alignment with antibody–antigen binding affinity. Moreover, antibody–antigen complexes constitute less than 3\% of the Protein Data Bank \cite{zhou2025comprehensive}, so affinity-relevant interface signals are vastly diluted during pre-training. As a result, even state-of-the-art unsupervised PLMs achieve only modest Spearman correlations ($\rho \approx 0.3\,\text{-}\,0.5$) with real in vitro affinity value on public benchmarks \cite{zhao2025benchmark}. 
This indicates that the current PLMs struggle to design antibodies with high affinity. Recently, some studies have introduced preference optimization into protein language models. However, most of these works are not trained for antibody binding affinity. One representative model is ProteinDPO \cite{widatalla2024aligning}, which fine-tunes ESM Inverse Folding (ESM-IF) \cite{hsu2022learning} on protein stability–annotated data using Direct Preference Optimization (DPO) \cite{rafailov2023direct}. However, ProteinDPO yields limited gains when transferred to affinity prediction tasks. Moreover, it updates all model parameters, which may make it difficult to retain the foundational model’s generalizable knowledge. In addition, DPO defines its implicit reward as the log-ratio between the policy model and a reference model, which has misalignment between training reward and inference metric, this will also affect the performance of preference learning \cite{meng2024simpo}. 

\begin{figure*}[t]      
  \centering
  \includegraphics[width=1\textwidth]{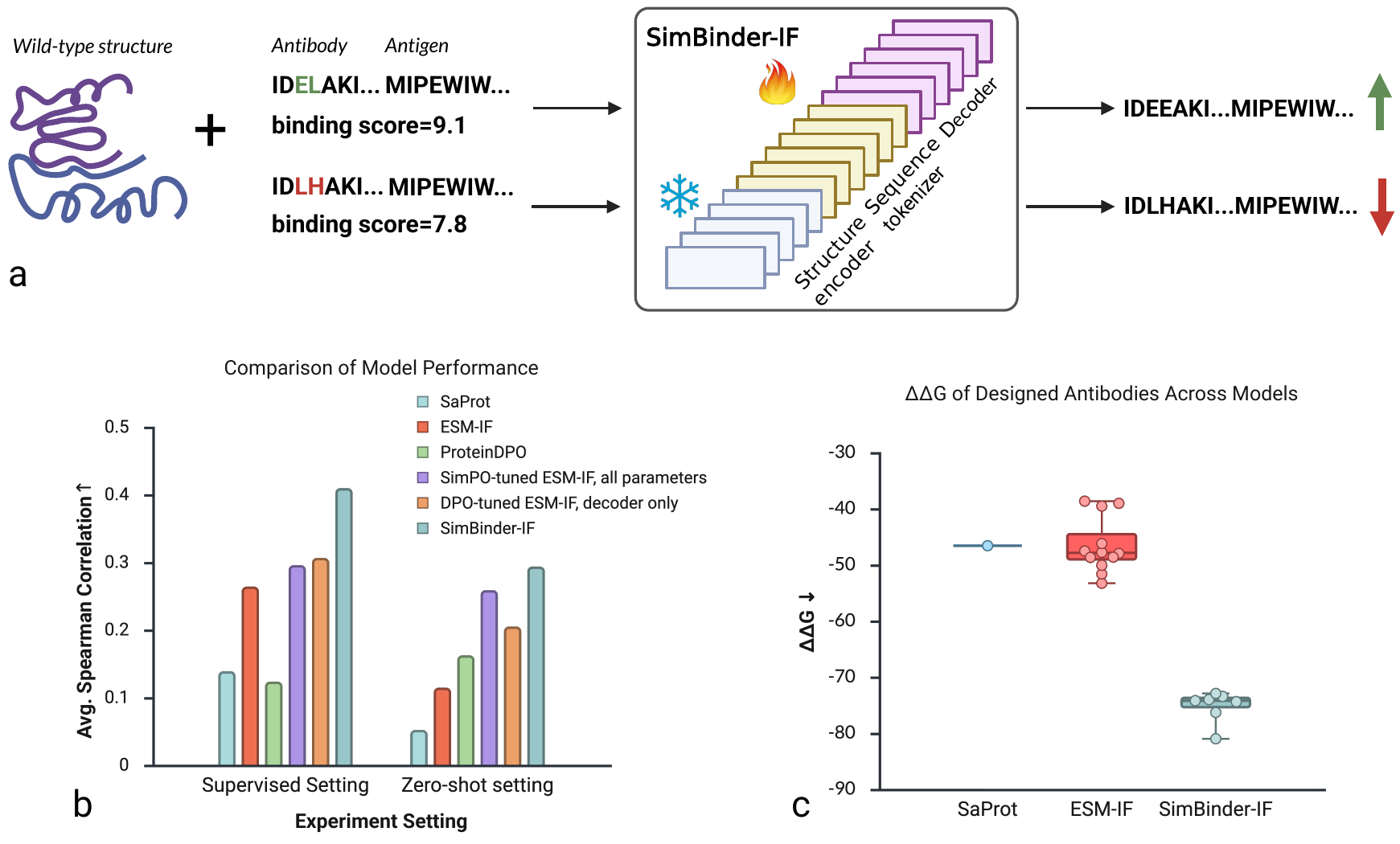}
  \caption{\textbf{Overview of the proposed SimBinder-IF and its benchmarking results.} (a) \textbf{SimBinder-IF Training Workflow.} Given antibody-antigen sequence pairs with associated 3D wild-type structures and binding scores, SimBinder-IF is trained using Simple Preference Optimization (SimPO) to assign higher log-likelihoods to sequences with better binding affinity. During this process, the structure encoder is frozen to preserve spatial understanding in the base model and enable efficient fine-tuning. (b) \textbf{Average Spearman correlation between model log-likelihood and binding affinity.} SimBinder-IF achieves state-of-the-art performance, as evidenced by a higher Spearman correlation between the model's log-likelihood scores and experimentally measured binding affinities. A higher correlation means the model better aligns its log likelihood scores with the experimentally measured binding affinities. These gains are consistently observed across different experiment settings, validating the generalizability and robustness of our framework. This figure encapsulates the two central innovations of the work: (1) a frozen structure-aware encoder-decoder design for efficient adaptation, and (2) effective application of preference-based optimization in protein sequence design tasks. (c) \textbf{The distribution of FoldX predicted binding free energy change} ($\Delta \Delta G$, lower is better) for antibodies generated by each model. $\Delta \Delta G$ measures the difference in $\Delta G$ between a mutant and the wild-type antibody. If the mutant has a more negative $\Delta G$ than the wild-type, it indicates stronger binding.}
\label{overview}
\end{figure*}

ESM-IF \cite{hsu2022learning} is a PLM originally designed for the inverse folding task, where the goal is to generate the protein sequence based on its structure. Given its state-of-the-art performance on antibody binding affinity tasks \cite{zhao2025benchmark}, we adopt ESM-IF as our base model and introduce SimBinder-IF, a conditional generative model that repurposes ESM-IF for antibody generation targeting improved binding affinity (Figure \ref{overview}a). SimBinder-IF is trained using Simple Preference Optimization (SimPO) \cite{meng2024simpo} on antibody binding affinity data by encouraging sequences with higher binding affinity to receive higher log-likelihood than those with lower affinity. In addition, to maintain the model’s ability to interpret protein structure, we keep the structure encoder of the base model fixed and update only the decoder during fine-tuning.

Our experiments on the AbBiBench benchmark \cite{zhao2025benchmark} validated the effectiveness of SimBinder-IF model. SimBinder-IF yields a 55\% relative improvement in average Spearman correlation across 11 antigen mutants between model predicted log-likelihood and in vitro affinity value in the supervised setting (from 0.264 to 0.410), and a 156\% relative improvement across 4 unseen antigens mutants (from 0.114 to 0.294), relative to the vanilla ESM-IF. Additionally, we evaluated the SimBinder-IF model in its ability to increase binding aﬀinity of anti-Influenza antibody F045-092 towards post-pandemic H1N1 antigens. Antibody mutants generated using the SimBinder-IF were ranked based on their predicted structural integrity and biophysical properties of the antibody–antigen complex. The fine-tuned model significantly outperforms other baselines in terms of predicted binding free energy ($\Delta \Delta G$; Figure \ref{overview}c). Moreover, antibodies generated by SimBinder-IF exhibit the best or comparable sequence plausibility and structural credibility compared to the other models tested (Figure \ref{casestudy}). These results show that our preference learning strategy markedly improves the base model's capacity to generate high-affinity antibody candidates with potential for therapeutic development. 

The remainder of this paper is organised as follows. Section \ref{results} summarises the benchmark performance and case-study findings; Section \ref{method} details the model architecture, training details and dataset; Section \ref{related_work} reviews current protein language models and preference-optimized PLMs, and examines log-likelihood as a proxy for antibody–antigen binding affinity; Section \ref{conclusion} concludes with limitations and future directions.

\section{Result and Discussion}
\label{results}

\subsection{SimBinder-IF Overview and Experimental Design} 
As illustrated in Figure \ref{overview}a, SimBinder-IF is an inverse folding architecture that takes a protein structure as input and generates the probability of each amino acid sequentially from left to right. To leverage the capabilities of a general protein model, the parameters of SimBinder-IF are initialized from ESM-IF, a state-of-the-art inverse folding model. During training, we freeze the parameters of the structure encoder and fine-tune only the decoder using Simple Preference Optimization (SimPO) \cite{meng2024simpo} on paired high- and low-affinity variants.
Training and evaluation rely on the 11-assay AbBiBench dataset ($\approx$155 k antibody–antigen mutants spanning viral and oncology targets); a detailed description of each assay appears in Table \ref{dataset}, \ref{sec_dataset} Experimental Dataset. Complete hyper-parameter and hardware settings are provided in \ref{train_detail} Training Details. The baselines compared in the experiments are introduced in \ref{baselines} Baselines.

\subsection{Binding Affinity Ranking}
\begin{figure}[htbp]
  \centering
  \includegraphics[width=0.5\textwidth]{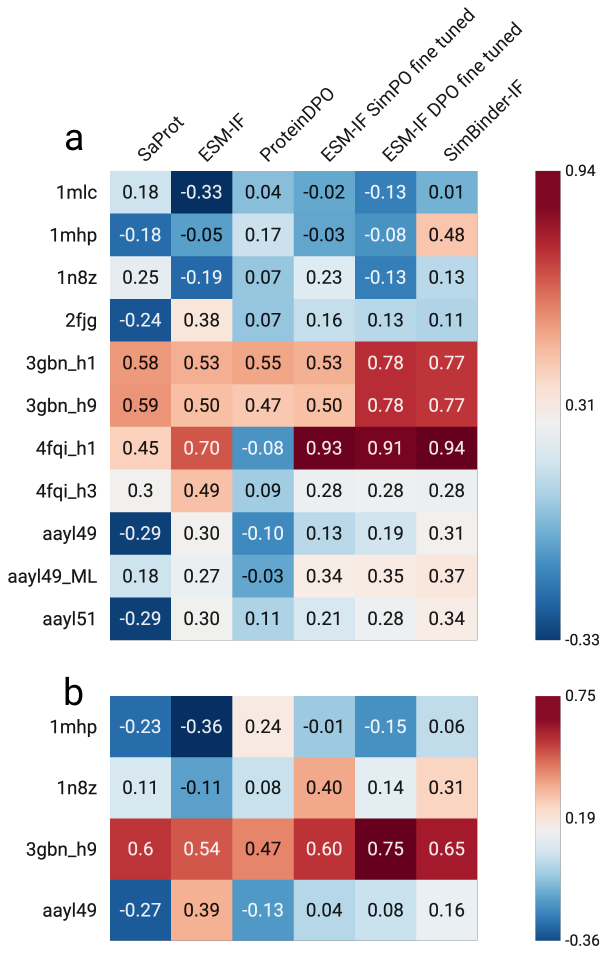} 
  \caption{Spearman correlation between predicted and experimentally measured binding affinity for antibody–antigen complexes across different models and datasets. (a) Supervised setting: Models are evaluated on antigens seen during training. (b) Zero-shot setting: Models are evaluated on held-out antigen–antibody pairs not used during training. Datasets include a diverse range of antigens, spanning both oncology and viral targets. Across both settings, decoder-only SimPO fine-tuning of ESM-IF consistently outperforms other models, demonstrating substantial gains in affinity prediction accuracy and generalization to unseen complexes.}
  \label{fig:heapmap}
\end{figure}

\begin{figure}[htbp]
  \centering
  \includegraphics[width=0.5\textwidth]{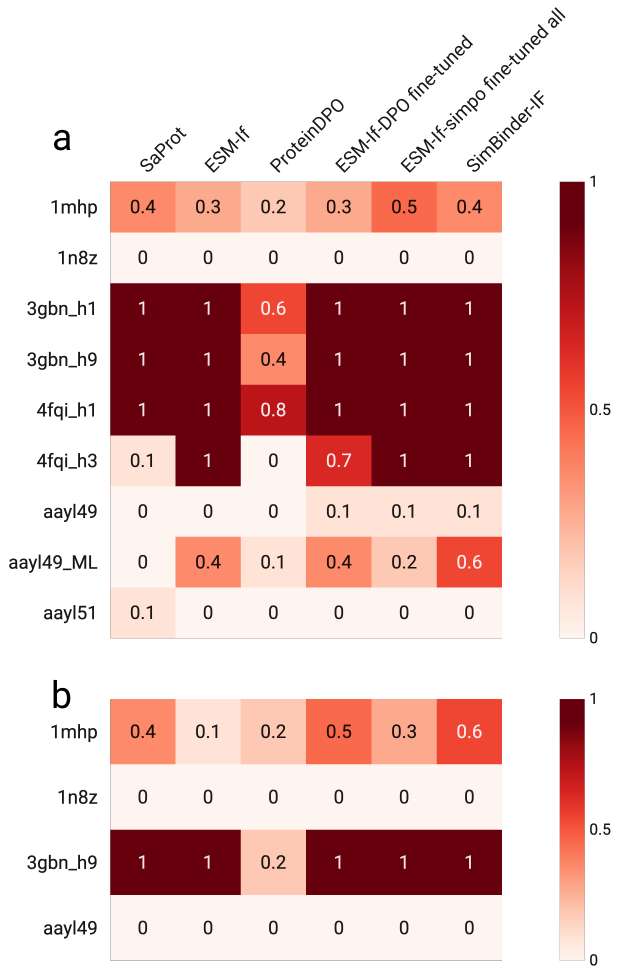} 
  \caption{Proportion of top-10 ranked antibody designs achieving  $\geq 10$-fold affinity improvement across models and datasets. The top and bottom parts show the results of the supervised and zero-shot settings, respectively. Only datasets reporting affinity as $-\log K_{d}$ were used. Datasets based on enrichment scores were excluded, as enrichment reflects relative sequence abundance and cannot determine fold change. In both evaluation settings, fine-tuned models achieved better performance than non-trained models, and among them, the ESM-IF model fine-tuned with SimPO by updating only the decoder parameters achieved the best results.}
  \label{fig:n_fold}
\end{figure}

We evaluated how preference fine-tuning affects ESM-IF’s capacity to predict antigen–antibody binding affinities in a \textbf{zero-shot setting} (held-out antigens). To measure how well a model’s zero-shot predicted log-likelihood aligns with wet-lab verified affinity, we calculated the Spearman correlation between the model likelihood and experimentally measured binding affinity. A high correlation indicates that the model assigns a higher likelihood to strong binders, suggesting that it can identify affinity-enhancing mutations in a zero-shot setting. To evaluate how effectively a model prioritizes the most promising antibodies, we also reported 10-fold precision@10 – the proportion of top 10 ranked variants that achieve at least 10-fold improvement
in binding affinity compared to the wild type. The fold change was calculated by taking the difference between the $pK_d$ values of the mutant and wild-type antibodies (i.e., $-\log_{10} K_d$), and exponentiating the result as $10^{\left(pK_{d,\;\text{mutant}} - pK_{d,\;\text{wild-type}}\right)}$. This yields the ratio of the dissociation constants $K_d$ between the wild-type and mutant antibodies. The Spearman's correlation coefficient score of different models across different antigens are shown in Figure \ref{fig:heapmap}, and the 10-fold improvement @10 is shown in Figure \ref{fig:n_fold}.


Decoder-only SimPO fine-tuning attained an average Spearman correlation of $0.294$, markedly surpassing the ESM-IF base ($\rho = 0.115$, $\Delta = +0.179$). This suggests that SimBinder-IF has greater potential than the base model in designing high-affinity antibodies. Importantly, the strong zero-shot gains demonstrate that SimBinder-IF substantially improves ESM-IF’s generalization to unseen antigens mutants.  

\subsection{Paratope Prediction Task}
\begin{figure}[htbp]
  \centering
  \includegraphics[width=0.5\textwidth]{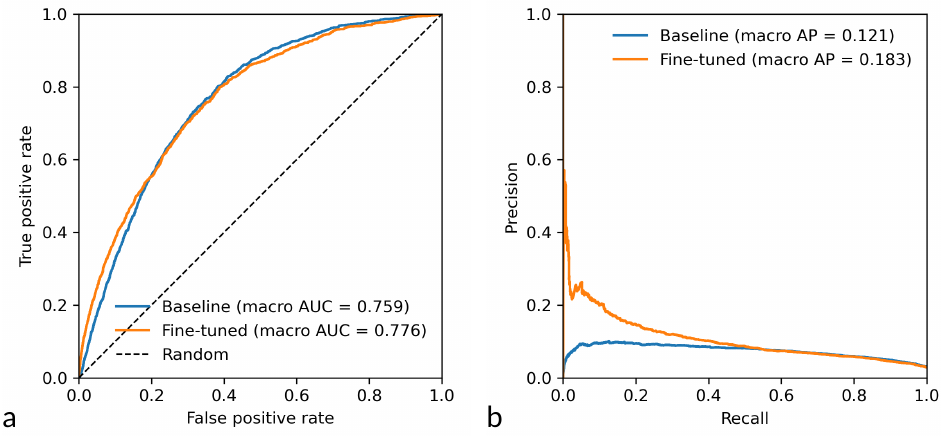} 
  \caption{Global paratope prediction performance comparison. (a) Global receiver operating characteristic (ROC) curves on the TDC Paratope test set. (b) Global precision–recall (PR) curves on the same test set, illustrating the trade‑off between precision and recall for the same set of models.}
  \label{fig:roc_pr_curves}
\end{figure}

To further investigate the antibody modeling capability of SimBinder-IF, we conducted paratope prediction experiments. A paratope, or antigen-binding site, is the antibody region that binds an epitope. Although we know the hypervariable regions drive binding, it remains challenging to pinpoint the interacting amino acids. This task predicts which antibody residues are positioned to bind the antigen.
Based on ESM-IF or SimBinder-IF, we design a simple architecture for this task. We freeze all base model parameters and train only a lightweight per-residue classification head on top of the residue-level representations. This head is implemented as a two-layer feedforward network that maps each residue embedding to a paratope probability.

We use the TDC Paratope dataset for our experiments, where antibodies are annotated with residue-level paratope labels. We obtain antibody structures from SAbDab, and antibody sequences are aligned to their corresponding structural chains so that the frozen base model can leverage the 3D structural context when producing token-level features for each antibody residue during training and evaluation.

Figure \ref{fig:roc_pr_curves} shows a comparison of global paratope prediction performance between ESM-IF and SimBinder-IF. Although SimBinder-IF yields a modest improvement in the area under the ROC curve (ROC AUC), it leads to a substantial gain in Precision–Recall performance, which is more informative for the highly imbalanced paratope prediction task. This indicates that fine-tuning primarily improves the base model’s ability to identify true paratope residues with higher precision, rather than merely increasing overall separability.

\subsection{Case Study: Redesign of the F045-092 Antibody to Target pdmH1N1 Influenza Hemagglutinin}
\begin{figure*}[t]      
  \centering
  \includegraphics[width=1\textwidth]{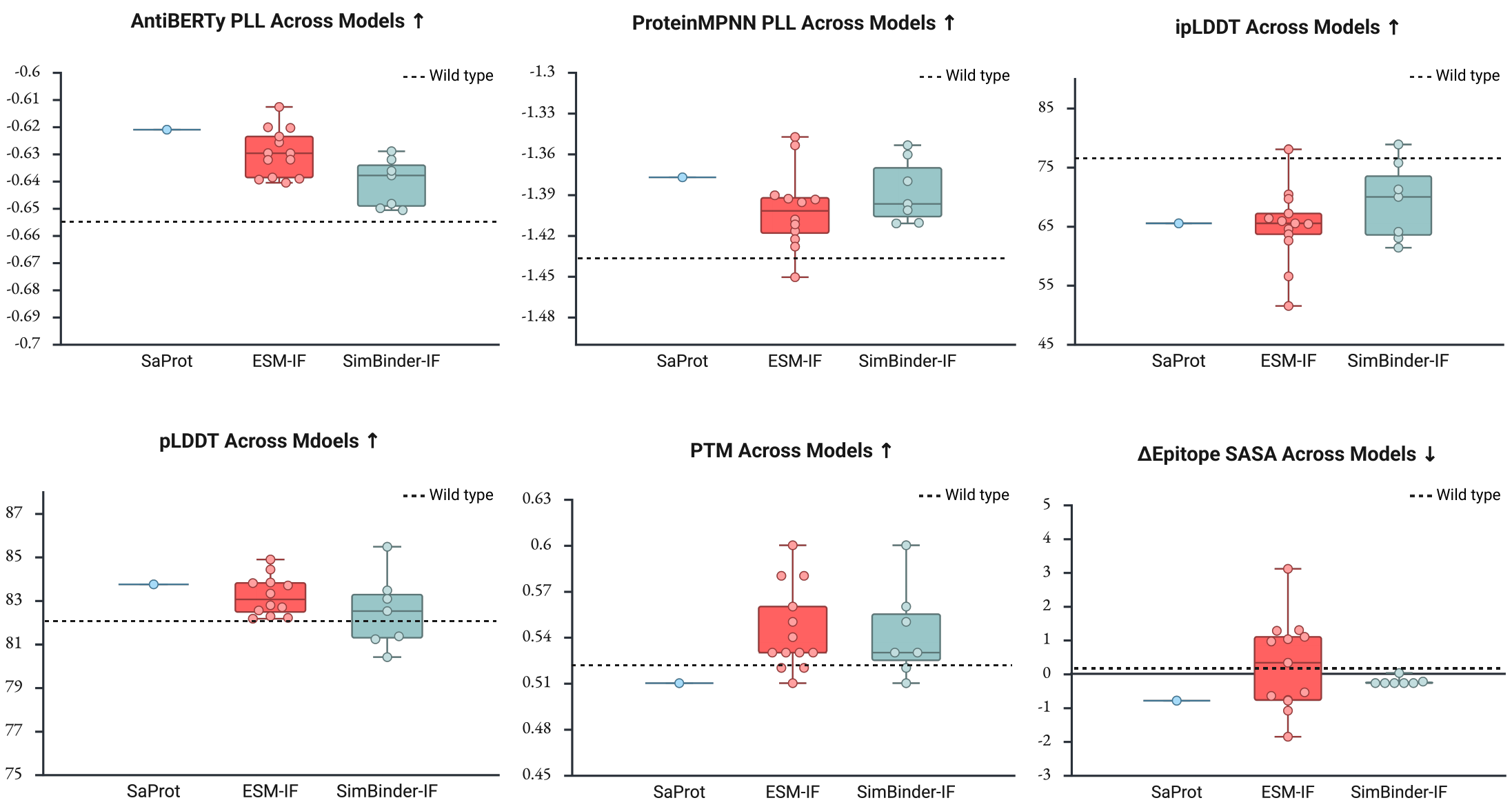}
  \caption{\textbf{Multi-metric evaluation of Pareto-optimal variants derived from the F045-092 antibody for pdmH1N1 binding activity.} The model used to generate each set of variants is indicated at the bottom of each panel. Six complementary metrics are displayed: PTM score, AntiBERTy PLL, ProteinMPNN PLL, pLDDT, interface-focused ipLDDT, and $\Delta$Epitope SASA. From the 1,500 variants generated by each model, evaluation across the six metrics yielded 20 Pareto-optimal candidates: 1 from SaProt, 12 from vanilla ESM-IF, and 7 from SimBinder-IF. Variants generated by SimBinder-IF (green) achieve the best in ProteinMPNN PLL, iPLDDT and PTM metrics, and near-best scores on the remaining three metrics. These results collectively demonstrate that SimBinder-IF yields antibody variants that are both sequence- and structure-reliable, with predicted higher binding affinity.}
\label{casestudy}
\end{figure*}
To assess the capacity of PLMs on their ability to generate antibody variants with improved binding affinity towards a specific antigen, we conducted a case study on anti-influenza A antibody F045-092 \cite{ohshima2011naturally}. This antibody showed little to no binding towards A/California/04/2009 (pdmH1N1) strain. Details of this antibody are provided in Section \ref{dataset_case_study}. Based on the AbBiBench results, we systematically evaluated the generative capacity of our SimBinder-IF model and compared it in parallel with ESM-IF and SaProt \cite{SuHZSZY24} on their ability to redesign the CDR-H3 region of F045-092. We generated over 1,500 variants per model under a constraint of up to five amino acid substitutions. All generated variants underwent a two-stage screening process where variants were first selected based on sequence plausibility (AntiBERTy \cite{ruffolo2021deciphering} likelihood) and predicted binding energy change ($\Delta \Delta G$). This initial pool was then refined using structure-based metrics, including complex structure confidence (pLDDT), Predicted Template Modeling score (PTM), epitope solvent-accessible surface area (SASA), and inverse folding likelihood (ProteinMPNN \cite{dauparas2022robust}). Details of the two-stage screening process and the introduction of metrics used are provided in Section \ref{two-stage}. Across the 1,500 in-silico designs generated per model, the two-stage screening pipeline distilled a Pareto-optimal panel of 20 variants: 1 from SaProt, 12 from the vanilla ESM-IF, and 7 from SimBinder-IF. Detailed analyses of variants in the Pareto set are presented in Figures \ref{overview} c and \ref{casestudy}. From these results, we have the following observations:

\begin{enumerate}
    \item \textbf{Binding energy change.} The SimBinder-IF variants exhibit markedly lower predicted binding energy, with a mean FoldX $\Delta \Delta G$ of $-75.16 kcal mol^{-1}$ versus $-46.57 kcal mol^{-1}$ for ESM-IF designs, confirming that preference fine-tuning translates into tighter antigen engagement.
    \item \textbf{Structure-aware quality metrics.} Relative to ESM-IF, SimBinder-IF variants score higher on ProteinMPNN log-likelihood, ipLDDT, and epitope $\Delta SASA$ , while matching both ESM-IF and SaProt on PTM and pLDDT confidence. These gains show that the improved binding affinity is not achieved at the expense of predicted foldability (pLDDT), global structural confidence (pTM), or interfacial packing quality (ipLDDT and $\Delta$Epitope SASA).
    \item \textbf{Sequence plausibility.} Although SimBinder-IF shows a marginally lower AntiBERTy PLL than ESM-IF, the difference is modest and does not offset the substantial advantages observed in structure-based metrics and $\Delta \Delta G$, suggesting that sequence plausibility remains within an acceptable range.
\end{enumerate}
SimBinder-IF consistently generates antibody variants that exhibit both higher binding affinity and structural reliability compared to those from ESM-IF and SaProt. 
\subsubsection{Binding Site Analysis of ESM-IF vs SimBinder-IF}
We used AlphaFold 3 to model ESM-IF- and SimBinder-IF-designed antibody variants complexed with the antigen. Based on the predicted structures, for the H1 hemagglutinin (HA) antigen, we observed a marked difference in the binding site distribution between antibodies designed by ESM‑IF and those by SimBinder‑IF. Among the ESM-IF-designed antibodies, 8 out of 12 primarily bind to the antigen’s head region, while the remaining bind to the middle region. In contrast, 6 out of 7 SimBinder-IF-designed antibodies bind to the head region, with only 1 primarily targeting the middle region. The head region of H1 HA constitutes the principal immunodominant site \cite{wu2020influenza}. The Figure \ref{casestudy} presents the structural superposition of the F045-09/pdmH1N1 complex, a representative ESM-IF–designed variant/pdmH1N1 complex, and a representative Binder-IF–designed variant/pdmH1N1 complex. This figure clearly illustrates the binding sites of all three antibodies on the same antigen.
\begin{figure}[t]      
  \centering
  \includegraphics[width=0.5\textwidth]{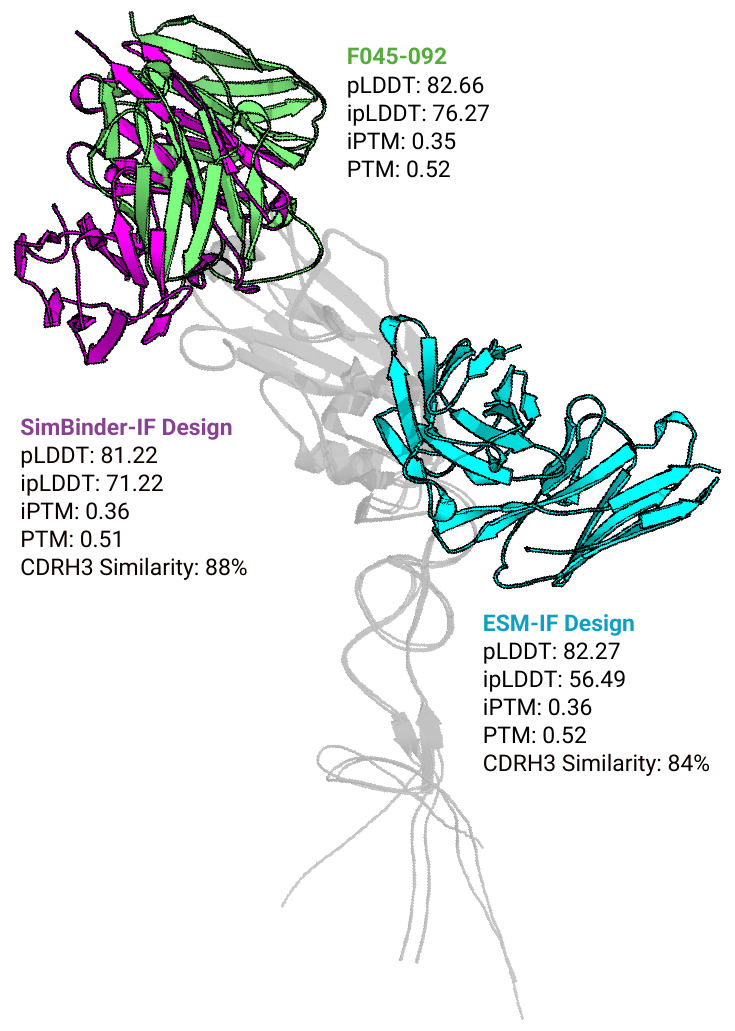}
  \caption{\textbf{Superposition of antibody variant–pdmH1N1 complexes.} The bright green structure represents the original F045-092 antibody. The bright red structure is a representative variant designed by SimBinder-IF that binds to the head region of pdmH1N1, while the bright blue structure represents a variant designed by ESM-IF that binds to the middle region. The antigen is shown in gray. All complexes were predicted using AlphaFold 3. Among the ESM-IF designs, 4 out of 12 primarily target the HA middle region, whereas the remaining variants bind to the head region. In contrast, SimBinder-IF variants more consistently engage the HA head region (6 out of 7), which is the major site for antibody recognition and development.}
\label{casestudy}
\end{figure}
\subsection{Biological implications and limitations}
The substantial improvements in Spearman correlation—55\% in the supervised setting and 156\% in the zero-shot setting—along with superior precision@10 compared to vanilla ESM-IF, demonstrate that SimBinder-IF aligns sequence likelihood with functional binding in a parameter-efficient manner. For an autoregressive language model, this suggests that SimBinder-IF holds greater potential than the base model for designing high-affinity antibodies. Notably, the strong zero-shot gains indicate that SimBinder-IF significantly enhances ESM-IF’s generalization to unseen antigen mutants. 

In the case study, variants designed by SimBinder-IF also outperform or match baseline designs in terms of ProteinMPNN likelihood, pLDDT, and interface-specific ipLDDT, achieving higher binding affinity without compromising structural reliability. Furthermore, SimBinder-IF-generated antibodies preferentially target the HA head region, a principal immunodominant site. Hemagglutinin is a class I viral fusion protein that forms a homotrimer, with each protomer comprising a membrane-distal globular head and a membrane-proximal stalk region. The head domain contains the receptor-binding site (RBS), which mediates viral attachment to host cells, and encompasses multiple antigenic sites \cite{wu2020influenza}. It is the principal target of antibodies elicited by seasonal influenza vaccines, and many neutralizing antibodies act by blocking the RBS or covering lateral patches on the head \cite{bangaru2019site}. However, some broadly neutralizing antibodies (bnAbs) recognize conserved epitopes at the interface between the head and stalk regions. These mid-region epitopes tend to be more conserved and can enable cross-reactive neutralization across different strains or subtypes over multiple years \cite{wu2020influenza}. Therefore, this head-localized binding preference suggests that SimBinder-IF tends to generate antibodies that mimic antibodies generated from natural immune responses, achieving potent neutralization by targeting the RBS. Collectively, these results underscore the practical utility of SimBinder-IF for efficient and effective antibody maturation against emerging influenza strains.

However, the head-centric bias of SimBinder-IF could restrict breadth, as mutations in the receptor-binding site accumulate rapidly; future iterations could add an on-the-fly epitope-diversification module to preserve this speed while gradually expanding coverage. A second limitation is the single-objective reward—binding affinity—leaving other developability attributes such as immunogenicity, viscosity, and manufacturability unaddressed. Moreover, current validation relies primarily on in-silico surrogates; limited wet-lab confirmation means real-world gains may diverge. Finally, freezing the structure encoder prevents adaptation to backbone rearrangements that destabilising mutations might induce, potentially capping performance when loop remodelling is essential. Consequently, integrating multi-objective SimPO, selectively unfreezing SE(3)-equivariant encoder layers, and coupling with high-throughput experimental feedback loops constitute promising avenues for expanding the therapeutic applicability of SimBinder-IF. 
\section{Methods and Materials}
\label{method}
\subsection{SimBinder-IF Overview}
As illustrated in Figure \ref{overview} a, we finetune a pretrained ESM inverse folding (ESM–IF) \cite{hsu2022learning} model by freezing its structure encoder and retraining only the sequence decoder. By freezing the encoder, we preserve its foundational ability to capture intricate structural features of antibody–antigen complexes, prevent catastrophic forgetting of pretrained knowledge, and reduce computational overhead. The model ingests the three dimensional structure of the wildtype antibody–antigen complex together with the sequence of a mutated antibody variant—where mutations are confined to CDR regions and concatenated with its antigen sequence—and outputs the log likelihood of that mutated antibody–antigen sequence conditioned on the structural context. To ensure that higher likelihoods correspond to stronger binding, we employ Simple Preference Optimization (SimPO) \cite{meng2024simpo}: during training, paired high and low affinity variants are presented simultaneously, and the decoder is optimized to increase the score of experimentally validated high affinity sequences while decreasing that of low affinity ones. The finetuned model demonstrates improved correlation with binding affinity in both zero-shot and supervised benchmarks and can generate novel CDR sequences predicted to enhance antigen binding. 

\subsection{Recap of the Base Model: ESM Inverse Folding}
The inverse folding problem involves inferring, from a fixed three‑dimensional backbone conformation, the amino acid sequence most likely to fold into and stabilize that structure.  In our previous work AbBiBench \cite{zhao2025benchmark}, we benchmarked various protein modeling approaches and find that inverse folding methods—including ProteinMPNN \cite{dauparas2022robust}, ESM‑IF, and AntiFold \cite{hoie2023antifold}—outperform other model classes in the antibody design task. ESM–IF, as a representative inverse‑folding model, is a graph‑based transformer trained to predict amino acid identities from 3D backbone coordinates. Its architecture comprises:

\begin{itemize}
  \item An invariant structural encoder \(f_{\mathrm{enc}}\) that embeds atomic positions and pairwise geometric features into residue‑level embeddings
  \[
    \mathbf{E} = f_{\mathrm{enc}}(X) \in \mathbb{R}^{n \times d},
    \quad
    X = \{\mathbf{x}_i\}_{i=1}^{3n}.
  \]
  \item A sequence tokenizer $f_{\text{tok}}$ maps each previously generated residue $y_{<i}$ into a learned embedding, which is concatenated with the structural embedding $\mathbf{E}$ to form the decoder input. An autoregressive decoder $f_{\text{dec}}$ predicts the next residue conditioned on both the structure and the partial sequence:
  \[
  \begin{aligned}
    p(Y \mid X) &= \prod_{i=1}^n p\bigl(y_i \mid y_{<i}, \mathbf{E}\bigr), \\
    p\bigl(y_i \mid y_{<i}, \mathbf{E}\bigr)
    &= \mathrm{softmax}\bigl(W\,h_i + b\bigr),
  \end{aligned}
  \]
  \[
    h_i = f_{\mathrm{dec}}(\mathbf{E}, y_{<i}) \in \mathbb{R}^d,
    \quad
    Y = (y_1, \dots, y_n).
  \]
\end{itemize}

During training, the model minimizes the negative log‑likelihood
\[
  \mathcal{L}_{\mathrm{NLL}}
  = -\sum_{(X,Y)} \log p(Y \mid X)
\]
to recover native sequences. 

Although originally designed for sequence recovery, we here reinterpret ESM–IF as a structure‑conditioned generative model: given the 3D coordinates of a wild‑type antibody–antigen complex, novel sequence variants can be sampled by drawing from \(p(Y \mid X)\) guided by the decoder’s log‑likelihood scores.

Freezing the GVP--GNN encoder and the Transformer encoder while training only the decoder reduces the number of trainable parameters to roughly $25$ M, about $18\%$ of the entire model, without losing the geometric conditioning provided by the frozen encoders.
\begin{figure}[t]      
  \centering
  \includegraphics[width=0.5\textwidth]{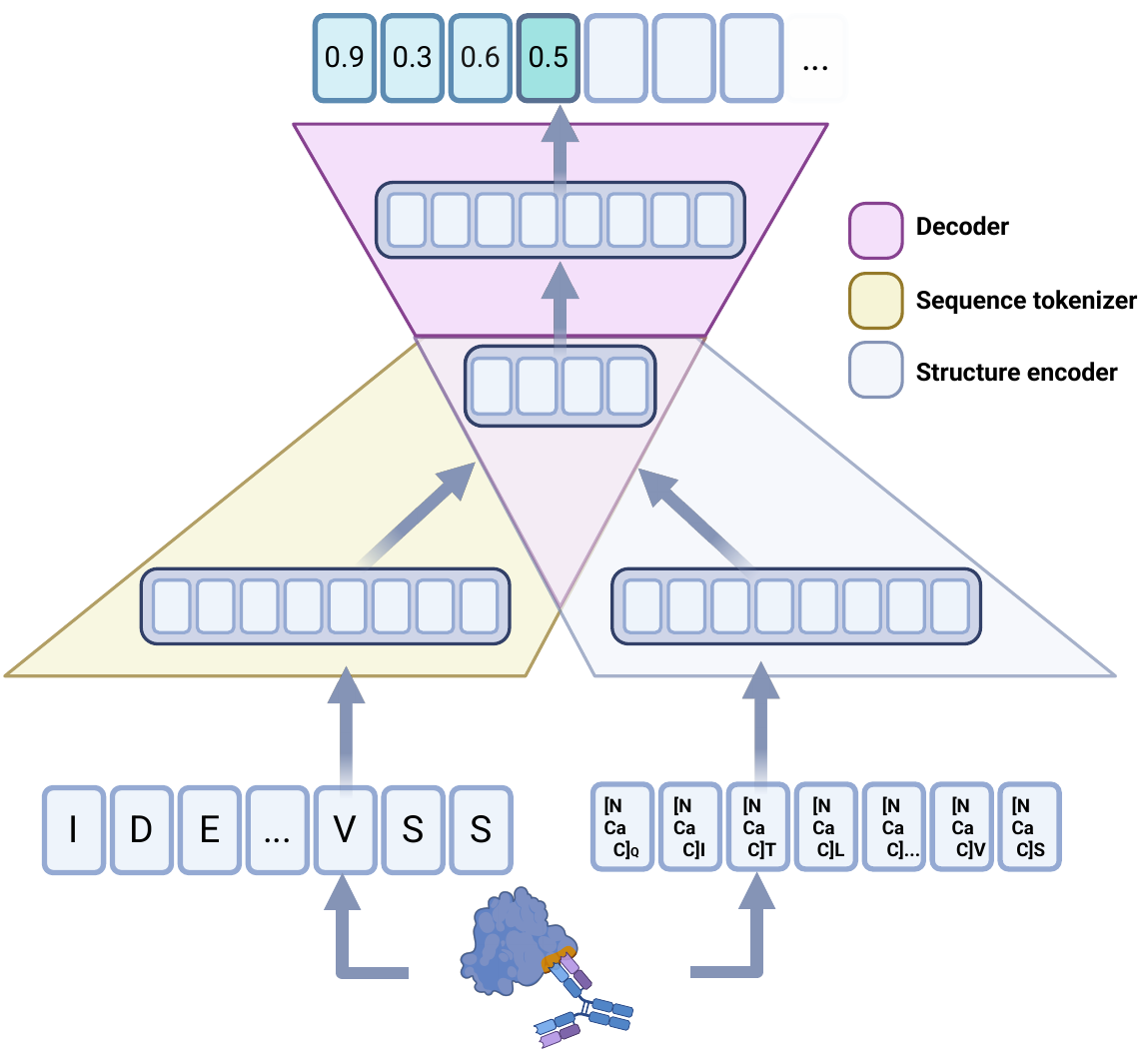}
  \caption{\textbf{Architecture of the SimBinder-IF.} SimBinder-IF builds on the inverse folding architecture of ESM-IF, comprising a structure encoder, a sequence tokenizer and a decoder. The structure encoder processes the 3D coordinates of an antibody–antigen complex to extract residue-level geometric embeddings, which are fed into an autoregressive decoder that predicts the probability of each amino acid sequentially.}
\label{esm-if}
\end{figure}
\subsection{Simple Preference Optimization for Affinity}
Our goal is to fine‑tune ESM‑IF so that it preferentially generates antibody sequences with high binding affinity: for a given antigen, antibody–antigen complexes that bind more tightly should receive higher average token‑wise log‑likelihoods, whereas lower‑affinity complexes should receive lower log‑likelihoods.

Fine‑tuning the decoder of ESM‑IF requires an objective that reflects how the model will be used at inference, namely, ranking candidate antibody sequences by their average token‑wise log‑likelihood. A natural starting point is Direct Preference Optimisation (DPO) \cite{rafailov2023direct}, which frames learning from pairwise preferences as a logistic‑ranking problem. For a prompt \(x\) and a sequence \(y\) the DPO reward is defined as
\[
r_{\mathrm{DPO}}(x,y)
=
\beta\bigl[\log \pi_{\theta}(y\mid x) \;-\;\log \pi_{\mathrm{ref}}(y\mid x)\bigr],
\]
where \(\pi_{\theta}\) is the current policy, \(\pi_{\mathrm{ref}}\) is a frozen reference model initialized from the unsupervised checkpoint, and \(\beta\) controls the reward scale. The loss over a preferred pair \((y_w,y_\ell)\) is obtained by maximizing the Bradley–Terry probability that the winner’s reward exceeds the loser’s:
\[
L_{\mathrm{DPO}}
=
-\log\sigma\bigl(r_{\mathrm{DPO}}(x,y_w)\;-\;r_{\mathrm{DPO}}(x,y_\ell)\bigr).
\]

While DPO is stable, it presents two limitations that become acute for structure‑conditioned antibody generation. First, the reward depends on the reference model, yet at inference only \(\pi_{\theta}\) is consulted. Consequently, a sequence can be favoured during training merely because the reference rates it even worse, even though its standalone likelihood remains low. This train–test mismatch is undesirable when absolute likelihood is the criterion used to rank variants in downstream pipelines. As shown in \cite{meng2024simpo}, in practice only \(\sim50\%\) of DPO‑trained triples preserve the desired likelihood ordering. Second, every optimization step requires a forward pass through the reference policy, doubling memory consumption and inflating wall‑clock time—a non‑trivial burden when fine‑tuning a large model like ESM‑IF on limited‑GPU hardware.

To resolve these issues we introduce SimPO, which removes the reference model entirely and aligns the optimization target with the metric employed at generation time. SimPO defines the reward as the length‑normalised log‑likelihood produced by the current policy,
\[
r_{\mathrm{SimPO}}(x,y)
=
\frac{\beta}{\lvert y\rvert}
\sum_{i=1}^{\lvert y\rvert}
\log \pi_{\theta}\bigl(y_i \mid x, y_{<i}\bigr),
\]
so the score used for learning is exactly the quantity used to rank sequences at test time. Incorporating this reward into the Bradley–Terry formulation and enforcing a fixed margin \(\gamma\) between the preferred and disfavoured sequences yields
\[
L_{\mathrm{SimPO}}
=
-\log\sigma\bigl(r_{\mathrm{SimPO}}(x,y_w)\;-\;r_{\mathrm{SimPO}}(x,y_\ell)\;-\;\gamma\bigr).
\]

Because SimPO no longer references an external policy, every gradient update directly raises the likelihood of preferred sequences, eliminating the spurious score inflation that can occur under DPO. The absence of the reference pass also cuts peak GPU memory and shortens each training epoch.

\subsection{Two-Stage Screening Pipeline for Novel Antibody Generation}
\label{two-stage}
This pipeline is proposed in AbBiBench as a method for evaluating the model’s ability to design novel antibodies \cite{zhao2025benchmark}. It begins with the protein model that redesign the CDR-H3 loop of F045-092 by introducing up to five point mutations. A residue position is included in the ``mutable pool'' only when the mutation’s log likelihood is higher than that of the wild type. Each model samples 1,500 candidate sequences while being conditioned on an AlphaFold-3 backbone of the predicted F045-092/H1N1 complex, yielding a diverse pool of variants.

Stage 1 – rapid prescreen. For every candidate, we calculate two fold-agnostic metrics: (i) the AntiBERTy log-likelihood \cite{ruffolo2021deciphering}, which quantifies evolutionary plausibility, and (ii) the FoldX-estimated binding free-energy change ($\Delta\Delta G$), which approximates binding potential. AntiBERTy is a transformer-based protein language model pre-trained on large antibody sequence dataset, and its log-likelihood score reflects the evolutionary plausibility of each antibody variant. $\Delta\Delta G=\Delta G^\text{mut} -\Delta G^{\text{wt}}$, where $\Delta G$ is the FoldX interaction free energy of the complex relative to the separated partners (kcal$\cdot$mol$^{-1}$); $\Delta\Delta G<0$ indicates a more favorable predicted interaction and thus higher affinity. Because $\Delta G\approx RT\ln K_d$, a decrease in $\Delta G$ corresponds to a lower predicted $K_d$ and stronger binding. Variants that rank within the top 20\% for both metrics advance to Stage 2.

Stage 2 – high-confidence ranking. The reduced subset is refolded as complete antibody–antigen complexes with AlphaFold 3. Three structure-based criteria are then evaluated: (i) complex-level pLDDT as a proxy for structural integrity, (ii) the change in epitope solvent-accessible surface area ($\Delta$SASA) to gauge interface burial, and (iii) the inverse-folding log-likelihood from ProteinMPNN to assess foldability. pLDDT is a per-residue confidence score on a 0–100 scale produced by AlphaFold that estimates local model accuracy under the lDDT-C$\alpha$ metric \cite{jumper2021highly}. Burial of hydrophobic residues—reflected by lower SASA—helps stabilize the protein’s hydrophobic core; therefore, if a variant exhibits a lower $\Delta SASA$ than the wild type, one can generally infer that the variant’s structure is more stable. ProteinMPNN conjoins a message-passing network \cite{gilmer2017neural} based backbone encoder and sequence decoder to iteratively recover amino acid sequences given the 3D coordinates of the protein backbone structure. If ProteinMPNN assigns a high inverse-folding likelihood, it judges that the mutated sequence is highly compatible with—and likely to fold well on—that AF3 predicted backbone. Final candidates were identified as the Pareto-optimal set across all evaluation criteria, providing a rigorous and biologically grounded novel antibody. Together with the AntiBERTy score and $\Delta \Delta G$ from Stage 1, these five metrics define a Pareto frontier; designs on this frontier are selected as the final high-affinity, structurally credible candidate mutants.

Besides above metrics, we also used PTM \cite{jumper2021highly,tunyasuvunakool2021highly} and ipLDDT when analysis the mutants in Pareto set. PTM is a global confidence metric in AlphaFold that estimates the similarity between the predicted model and the unknown true structure in terms of overall fold, weighted by per-residue resolvability.

\subsection{Training Details}
\label{train_detail}
\begin{figure}[htbp]
  \centering
  \includegraphics[width=0.5\textwidth]{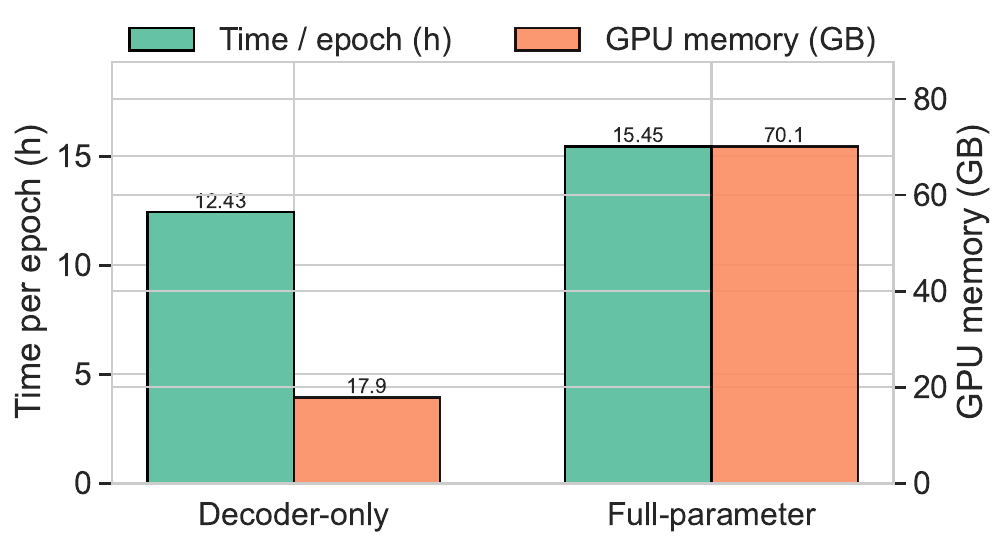} 
  \caption{Training efficiency comparison between decoder-only and full-parameter SimPO fine-tuning variants in terms of training time and GPU memory usage.}
  \label{fig:runtime_memory}
\end{figure}

We used the \texttt{AdamW} optimizer with a learning rate of $1\times10^{-4}$ and trained for 3 epochs using a batch size of 32. For SimPO loss, we set the temperature $\beta = 0.1$ and the margin $\gamma = 0.1$. In our experiments with DPO, the temperature was also set to $\beta = 0.1$. Model performance was evaluated using the Spearman correlation between predicted and actual binding scores, computed per antigen. The best-performing model on the validation set was saved and used for final evaluation on the test set. All training and inference were performed on a single NVIDIA H200 GPU with 141\,GB of memory. The measured wall-clock training times and peak memory usage for a batch size of 8 are shown in Figure \ref{fig:runtime_memory}. These results highlight the computational efficiency of our decoder-only strategy.

\subsection{Baselines} \label{baselines}
We establish baselines with 3 structure-based methodologies: SaProt \cite{SuHZSZY24}, ESM-IF, and ProteinDPO \cite{widatalla2024aligning}. SaProt is a masked language model for proteins that is also structure-aware. SaProt achieves leading performance in the AbBiBench test. ProteinDPO builds upon ESM-IF by fine-tuning the model via direct preference optimization for stabilizing variants over destabilizing variants. Besides vanilla ESM-IF, to ensure a comprehensive evaluation, we also include two ESM-IF variants trained on the AbBiBench dataset: ESM-IF fine-tuned with DPO while updating only the decoder parameters, and ESM-IF fine-tuned with SimPO across all model parameters.

\subsection{Experimental Dataset} \label{sec_dataset}
\subsubsection{Antibody Binding Affinity Ranking}
To train and test the efficacy of the proposed method, we utilize the AbBiBench dataset, which includes compiled data for 11 antibody-antigen binding-affinity assays (Table \ref{dataset}). Each assay includes curated data on the heavy chain sequence, experimental binding affinity measurements, and wild-type antibody-antigen complex structure. In total, 155,853 mutants are represented, and each assay ranges from 40 to 65,535 mutants. Moreover, experimental binding affinities were standardized from Kd values (dissociation constants) and enrichment ratios to binding scores via transformations using negative log and log, respectively. In turn, a higher binding score denotes more favorable binding across the entire dataset. Evaluation on AbBiBench is stratified according to each antigen-antibody assay to ensure a fair, accurate assessment over a range of structures.  

We appraise the performance of the model on 2 different settings: a \textit{supervised setting} and a \textit{zero-shot setting}. In the former, each antigen-antibody set is randomly arranged in a 60-30-10 train-test-validation split, and evaluation is performed on the test set after fine-tuning via the train and validation sets. On the other hand, we reserved four antigen–antibody mutation datasets as our test set—trastuzumab–HER2 (\texttt{1n8z}), AQC2–integrin-$\alpha$1 (\texttt{1mhp}), CR6261–influenza H9 hemagglutinin (\texttt{3gbn\_h9}), and AAYL49–SARS-CoV-2 Spike HR2 (\texttt{aayl49})—to ensure broad, biologically meaningful evaluation:

\begin{itemize}
      \item \texttt{1n8z} (trastuzumab–HER2) probes model performance on a clinically relevant oncological epitope.
      \item \texttt{1mhp} (AQC2–integrin-$\alpha$1) provides a structurally distinct integrin system, testing consistency across cancer-associated antigens.
      \item \texttt{3gbn\_h9} (CR6261–Influenza A/H9N2) challenges the model to capture mutational effects in a rapidly evolving viral protein.
      \item \texttt{aayl49} (AAYL49–SARS-CoV-2 Spike HR2) evaluates generalization from machine learning–generated phage-display libraries (aaly49\_ML) to experimentally measured viral binding data.
\end{itemize}

By spanning both oncology and viral contexts, and encompassing mutation landscapes from both computational and experimental sources, this selection provides a rigorous, diverse benchmark for assessing the generalization of fine-tuned protein language models.

\begin{table*}[ht]
\centering
\begin{tabular}{lllll}
\hline
\textbf{ID} & \textbf{Antibody} & \textbf{Antigen} & \textbf{Supervised Test Set Size} & \textbf{Zero-shot Test Set Size} \\
\hline
4fqi\_h1    & CR9114        & Influenza A/ New Caledonia/20/99 (H1N1)               & 19528 & --   \\
4fqi\_h3    & CR9114        & Influenza A/ Wisconsin/67/2005 (H3N2)                 & 19660 & --   \\
3gbn\_h1    & CR6261        & Influenza A/ New Caledonia/20/99 (H1N1)                &   566 & --   \\
3gbn\_h9    & CR6261        & Influenza A/ Hong Kong/1073/1999 (H9N2)                &   552 & 1842 \\
aayl49      & AAYL49        & Spike HR2                                              &  1294 & --   \\
aayl49\_ML  & AAYL49\_ML    & Spike HR2                                               &  2686 & 4312 \\
aayl51      & AAYL51        & Spike HR2                                               &  1296 & --   \\
2fjg        & G6.31         & VEGF                                                    &   667 & --   \\
1mlc        & D44.1         & Hen-egg-white lysozyme                                  &   369 & --   \\
1n8z        & trastuzumab   & HER2                                                    &   126 &  419 \\
1mhp        & AQC2          & Integrin-$\alpha$-1                                     &    12 &   40 \\
\hline
\end{tabular}
\caption{Overview of the 11 Ab-Ag binding-affinity assays in AbBiBench \cite{zhao2025benchmark} and their corresponding test set sizes.}
\label{dataset}
\end{table*}

\subsubsection{Case Study}
\label{dataset_case_study}
To evaluate the capacity of protein language models to design antibody variants with enhanced binding toward a given antigen, we performed a case study centered on F045-092 \cite{ohshima2011naturally}. This antibody was discovered through comprehensive screening of a human B-cell repertoire using phage-display technology. While initially raised against H3 influenza virus particles, this antibody also demonstrated a broad range of binding activity against different strains belonging to H1N1, H2N2 and H5N1 viruses \cite{ohshima2011naturally}. However, F045-092 shows little to no binding towards A/California/04/2009 (pdmH1N1) strain. This has been attributed to a steric bulge present in all H1N1 strains that emerged after 2009, caused by the insertion of a lysine residue at position 133a \cite{ekiert2012cross,simmons2023new}.  No experimentally solved structure of the F045/pdmH1N1 complex has been reported to date, and this antibody–antigen pair has not been previously explored in computational antibody design literature. Therefore, our study constitutes a fully novel evaluation setting, ensuring the absence of training–test data leakage. The structure of the F045/pdmH1N1 complex used in this study was generated using AlphaFold3.

\section{Related Work}
\label{related_work}

\subsection{Protein Language Models and Inverse Folding}
\label{related_work_plm}
Large-scale protein language models (PLMs) trained with self-supervision on hundreds of millions of natural sequences have transformed structure prediction and design. Sequence-only models such as ESM-2 \cite{hayes2025simulating}, SaProt \cite{SuHZSZY24}, and ProtGPT2 \cite{ferruz2022protgpt2} learn contextual residue embeddings that implicitly encode aspects of the folding landscape. Structure-conditioned inverse-folding models advance this idea by taking a fixed backbone as input and predicting sequences likely to adopt that conformation. ESM-IF couples a graph neural network encoder over atomic coordinates with an autoregressive decoder over residues and currently achieves state-of-the-art perplexity on CATH test folds and AbBiBench zero-shot test \cite{zhao2025benchmark}. Although such models generalize well across protein families, antibody design presents unique challenges—combinatorial CDR diversity, rigid–flexible interface dynamics, and the scarcity of high-quality paired antibody–antigen structures—which motivate task-specific adaptation strategies. 

\subsection{Log-Likelihood as a Proxy for Antibody–Antigen Binding Affinity}

Prior efforts in protein fitness benchmarking have demonstrated that model likelihood—or, equivalently, lower perplexity—correlates with functional performance. For example, the ProteinGym \cite{notin2023proteingym} benchmark measures the correlations between model perplexity and general protein fitness. FLAb \cite{chungyoun2024flab} extends this framework to antibodies by compiling experimental measurements of affinity, specificity, immunogenicity, and developability.  

Shanker et al \cite{shanker2024unsupervised} compute the joint log-likelihood of antibody sequences given their antigen-bound backbone using ESM-IF and shows that this score correlates strongly with experimentally determined binding affinities across five deep mutational scanning datasets (Spearman $\rho$ up to 0.65) . Building on these observations, AbBiBench \cite{zhao2025benchmark} explicitly treats the Ab–Ag complex as the input for zero‐shot affinity prediction. It computes the Spearman correlation between a model’s log-likelihood on the Ab–Ag complex and experimentally measured binding affinities. A consistently high correlation indicates that variants with stronger wet-lab affinities receive higher model likelihoods, thereby further underscoring the rationale for using log-likelihood to predict affinity rankings. 

\subsection{Existing Preference-Optimized Protein Models}
\label{related_work_po}
ProteinDPO \cite{widatalla2024aligning} fine-tuned ESM-IF with stability-based fitness pairs and improved enzyme thermostabilization rates; however, its objective centers on global stability rather than binding affinity, and it updates all parameters, risking the erosion of structure-aware priors. Multi-Objective Binder DPO \cite{mistani2024preference} optimized ProtGPT2 for receptor–peptide binding under multiple kinetic thresholds, yet the sequence-only architecture lacks explicit structural context needed for antibody–antigen interfaces. AbDPO \cite{zhou2024antigen} formulates antibody design as a preference‐optimization problem, using biophysical energy (total energy and binding energy) as the preference signal and employing Direct Preference Optimization. ResiDPO \cite{xue2025improving} applied preference learning to improve foldability using LigandMPNN with AlphaFold pLDDT as a signal—again focusing on collapsibility rather than affinity. CtrlProt \cite{liu2025controllable} introduces multi-attribute preference tuning on ProtGPT2 for controllable protein generation but does not consider antibody-specific attributes or exploit 3D coordinate information. 

In summary, although previous DPO-based approaches have advanced aspects of protein design, they share three key limitations: first, there remains a lack of models directly aligned with true antibody-antigen complex binding affinity; second, by scoring variants using the log-likelihood ratio between a fine-tuned model and reference model, DPO introduces a discrepancy between the training objective and the inference process; and third, most existing methods train the foundation model’s entire parameter set, which may result in the model losing its fundamental protein-modeling capabilities.

\section{Conclusion}
\label{conclusion}
By freezing the structure encoder and fine-tuning only the decoder with Simple Preference Optimization (SimPO), this work reframes ESM-IF as a structure-conditioned generator that delivers sizeable gains in binding-affinity prediction: average Spearman correlation rises from 0.264 to 0.410 across eleven antigens in the fully supervised setting, while zero-shot performance on four unseen complexes improves from 0.115 to 0.294—relative increases of 55\% and 156\%, respectively.

The study shows that such improvements do not require heavy parameter updates. Keeping the encoder fixed preserves structural priors and curbs over-fitting, yet the resulting lightweight model still outperforms both the untuned SOTA baseline and full-parameter alternatives; SimPO’s  reference-free reward further closes the gap between training and inference and carries over to unseen antigens. Together these findings suggest that structure-aware, parameter-efficient preference learning can surpass more cumbersome approaches while retaining flexibility for generative design.


Future work can extend SimPO to multi-objective preference learning that balances affinity with developability, and allow selective unfreezing or introduce SE(3)-equivariant modules so the structure encoder can accommodate mutation-induced conformational changes. Coupling the model with automated high-throughput screening would enable active-learning loops in which experimental feedback continually refines the reward, moving toward a comprehensive, data-driven platform for therapeutic antibody discovery.




\bibliographystyle{plain}
\bibliography{reference}

\end{document}